\documentstyle[makeidx,epsfig,wrapfig,amsbsy,amsfonts]{article}
\advance \topmargin by -\headheight
\advance \topmargin by -\headsep
\evensidemargin \oddsidemargin
\begin{document}
\pagestyle{plain}
\begin{titlepage}
\flushright{2002-xxx}
%\flushright{draft}
\flushright{\today}
\flushright{OEF}
%\prepnum{2002-xxx}{OEP }
\vspace*{0.15cm}
\begin{center}
{\Large\bf
    First results on a search for light pseudoscalar sgoldstino in
  $K^{-}   $ decays  
    }
\vspace*{0.15cm}

%\vspace*{0.5cm}
\vspace*{0.3cm}
{\bf  
I.V.~Ajinenko, S.A.~Akimenko,  
I.G.~Britvich, G.I.~Britvich, K.V.~Datsko,  A.P.~Filin, 
A.V.~Inyakin,  A.S.~Konstantinov, V.F.~Konstantinov, 
I.Y.~Korolkov,  V.M.~Leontiev, V.P.~Novikov,
V.F.~Obraztsov,  V.A.~Polyakov, V.I.~Romanovsky, 
  V.I.~Shelikhov, N.E.~Smirnov,   
  O.G.~Tchikilev,  E.V.~Vlasov,   O.P.~Yushchenko. }
  
\vskip 0.15cm
{\large\bf $Institute~for~High~Energy~Physics,~Protvino,~Russia$}

%\vskip 0.2cm
\vskip 0.35cm
{\bf 
 V.N.~Bolotov, S.V.~Laptev, V.A.~Lebedev,  A.R.~Pastsjak, A.Yu.~Polyarush,  
  R.Kh.~Sirodeev.}
\vskip 0.15cm
{\large\bf $Institute~for~Nuclear~Research~Moscow,~Russia$}
\vskip 0.15cm
\end{center}
\end{titlepage}
\begin{center}
Abstract
\end{center}
 A search for the light pseudoscalar sgoldstino production in three-body
 $K^{-}$ decay  $K^{-} \rightarrow \pi^{-}\pi^{0}~ P $ has been 
 performed with the ``ISTRA+'' detector exposed to the 25 GeV negative 
 secondary beam of the U-70 proton synchrotron. No signal is seen.  
 Upper limits for the branching ratio
 $Br (K^{-} \rightarrow \pi^{-} \pi^{0} P)$ 
at $90 \%$ confidence level vary between
 $2.0~ \cdot ~10^{-5}$ and $0.5~ \cdot ~10^{-5}$  in the
 effective mass $m_{P}$ range from 0 up to 190~MeV. Our results 
improve the limits published by the E787 Collaboration in the
 mass interval between 0 and 120~MeV and are the first ones at higher
 masses.

\raggedbottom
\sloppy

\section{ Introduction}

 In supersymmetric models with spontaneous supersymmetry breaking 
 superpartners of a Goldstone fermion --- goldstino,
 pseudoscalar $P$ and scalar $S$, should exist. In 
 various versions of gravity mediated and gauge mediated theories  
 one or both of these
 weakly interacting bosons --- sgoldstinos --- are light and therefore can
 be observed in kaon decays. Moreover, if sgoldstino interactions with quarks
 conserve parity,(as in left-right extensions of MSSM), and P is lighter than
 S, so that $m_{S}>(m_{K}-m_{\pi})$ and $m_{P}<(m_{K}-2m_{\pi})$, sgoldstinos 
 show up in the decay $K \rightarrow \pi \pi P$, rather than 
 in  much better constrained $K \rightarrow \pi S$.
 The phenomenology of light sgoldstinos in this scenario is considered in
 details in \cite{ref1}. 
 Under 
 assumption  that sgoldstino interactions with quarks and
 gluons violate quark flavor and conserve parity, low energy interactions
 of pseudoscalar sgoldstino $P$ with quarks are described by the Lagrangian:
\begin{equation}
 L = -P \cdot (h^D_{ij} \cdot \overline{d}_i i \gamma^5 d_j +
 h^U_{ij} \cdot \overline{u}_i i \gamma^5 u_j)
\end{equation}
 with
\begin{equation}
  d_i = (d,s,b)~, ~~~~~~ u_i = (u,c,t)~,
\end{equation}
  with coupling constants $h_{ij}$ proportional to the left-right soft
 terms in the matrix of squared masses of squarks:
\begin{equation}
   h^{D}_{ij} = \frac{\mbox{\~{m}}^{(LR)2}_{D,ij}}{\sqrt{2} F}~, ~~~~~~~~
   h^{U}_{ij} = \frac{\mbox{\~{m}}^{(LR)2}_{U,ij}}{\sqrt{2} F}~,
\end{equation}
 Here the scale of supersymmetry breaking is denoted as $\sqrt{F}$.  
  The constraints on the flavor-violating coupling of sgoldstino
  to quarks  is evaluated using the $K^{\circ}_L - K^{\circ}_S $ 
mass difference  and $CP$ violating parameter $\epsilon$ in the neutral kaon 
system: 
 $|h^{D}_{12}| \leq 7 \cdot 10^{-8};
 |Reh^{D}_{12} \cdot Imh^{D}_{12}| < 1.5 \cdot 10^{-17} $ .
 It has been shown in \cite{ref1}
   that, depending on the phase of sgoldstino-quark coupling, these
constraints result in the following limits on the branching ratio:
$Br (K^{-}\rightarrow \pi^{-} \pi^{0} P ) \leq 1.5 \times 10^{-6} 
\div 4\times 10^{-4}$ 
 (see Fig.~1 for the  diagram of the $K^-$ decay). A search for P in 
 charged kaon decays is of particular interest for the case 
 $Reh^{D}_{12} \sim 0$, when the corresponding branching ratio of $K_{L}$ is
 small. \\
 Light sgoldstino decays into two photons or into a pair of charged leptons,
 two photon decay dominating almost everywhere in the parameter space. 
 Depending on the parameter $g_{\gamma}=
 \frac{1}{2\sqrt{2}}\frac{M_{\gamma \gamma}}{F}$, where $M_{\gamma \gamma}$
 is the photino mass, sgoldstino decays either inside
 or outside the detector. In the present search we assume that sgoldstino decays
 outside the detector, i.e is invisible.
  The existing limits on the branching $Br (K^{-}\rightarrow \pi^{-} \pi^{0} P)$
are at the
level of $ 4~ \cdot 10^{-5}$ \cite{ref6}, whereas the limits for the
scalar sgoldstino $S$  can be estimated from the
studies of the $K^+ \rightarrow \pi^+ \nu \overline{\nu}$~ at the level of 
 $4.7 \cdot 10^{-9}$, see \cite{ref0} for  recent review.
\begin{figure}[H]
\epsfig{file=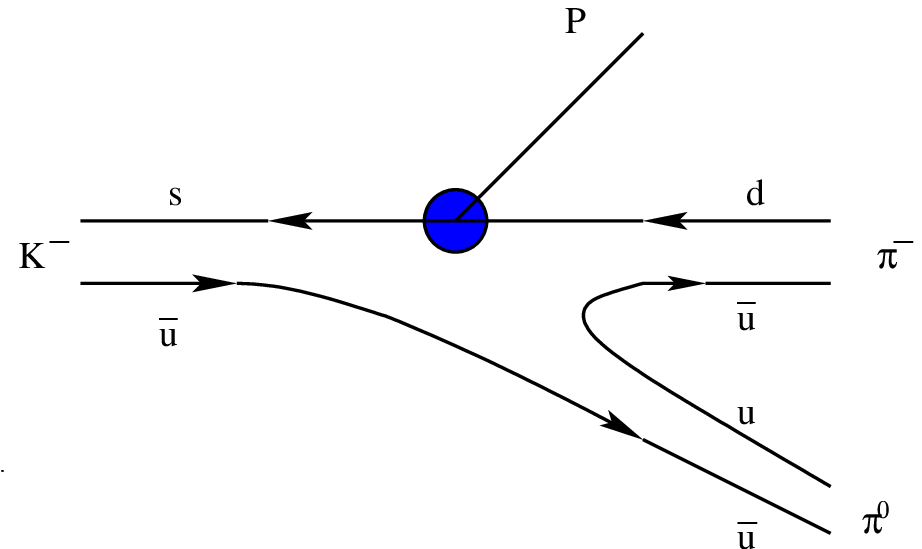,width=16cm}
\caption{$K^-$ decay into sgoldstino and pions.}
\end{figure} 
   
  The aim of our present   study  
 is to search for invisible pseudoscalar sgoldstino in the decay
 of $K^{-}$. Experimental setup and event selection are described
 in Section~2, the results of the analysis are presented in Section~3
 and the conclusions are given in the last Section.
 
\section{ Experimental setup and event selection}

\subsection{ Experimental setup}

The experiment is performed at the IHEP 70 GeV proton synchrotron U-70.
The  spectrometer  ISTRA+ has been described in some details in our
recent papers on $K_{e3}$  \cite{ref2}, $K_{\mu 3}$ \cite{ref3} 
  and $\pi^-\pi^{\circ}\pi^{\circ}$ decays \cite{ref4},
here we briefly recall characteristics relevant to our analysis. 
 The 40~m spectrometer ISTRA+ is located in the negative unseparated 
 secondary beam line 4A of U-70. The beam momentum  is $\sim 25$ GeV with 
$\Delta p/p \sim 2 \%$. The admixture of $K^{-}$ in the beam is $\sim 3 \%$,
 the beam intensity is $\sim 3 \cdot 10^{6}$ per 1.9 sec U-70 spill.
 A schematic view of the ISTRA+ setup is shown
 in Fig.~2. The momentum of a beam particle deflected by the magnet M$_1$
 is measured by four proportional chambers BPC$_1$---BPC$_4$ with 1~mm
 wire step, the kaon identification is done by three threshold Cerenkov
 counters \v{C}$_1$---\v{C}$_3$. The 9~meter long vacuumated decay
 volume is surrounded by eight lead glass rings used to veto low energy
 photons. The same role is played by 72-cell lead glass calorimeter SP$_2$.
 The decay products deflected in the magnet M$_2$ with 1~Tm field integral
  are measured with 2~mm step proportional chambers PC$_1$---PC$_3$, with
 1~cm cell drift chambers DC$_1$---DC$_3$ and finally with 2~cm diameter
 drift tubes DT$_1$---DT$_4$. A wide aperture threshold Cerenkov counter
 \v{C}$_4$ is filled with He and used to trigger electrons. SP$_1$ is a
 576-cell lead glass calorimeter, followed by HC --- a scintillator-iron
 sampling hadron calorimeter. MH is a 11x11 cell scintillating hodoscope,
 used to solve ambiguity for multitrack events and improve the time
 resolution of the tracking system, MuH is a 7x7 cell muon hodoscope.
 
  The trigger is provided  by scintillation counters S$_1$---$S_5$,
  beam Cerenkov counters and by the analog sum of amplitudes from
  last dinodes of the SP$_1$ : 
   T=$S_1 \cdot S_2 \cdot S_3 \cdot \overline{S}_4\cdot \mbox{\v{C}}_1
   \cdot \overline{\mbox{\v{C}}}_2 \cdot 
   \overline{\mbox{\v{C}}}_3 \cdot
   \overline{S}_5 \cdot \Sigma(SP_1)$,
   here S$_4$ is a scintillation counter with a hole
   to suppress beam halo, S$_5$ is a counter downstream of the setup at the
   beam focus, $\Sigma(SP_1)$ - a requirement for the analog sum to be
   larger than a MIP signal.
 ,
\begin{figure}[H] 
\epsfig{file=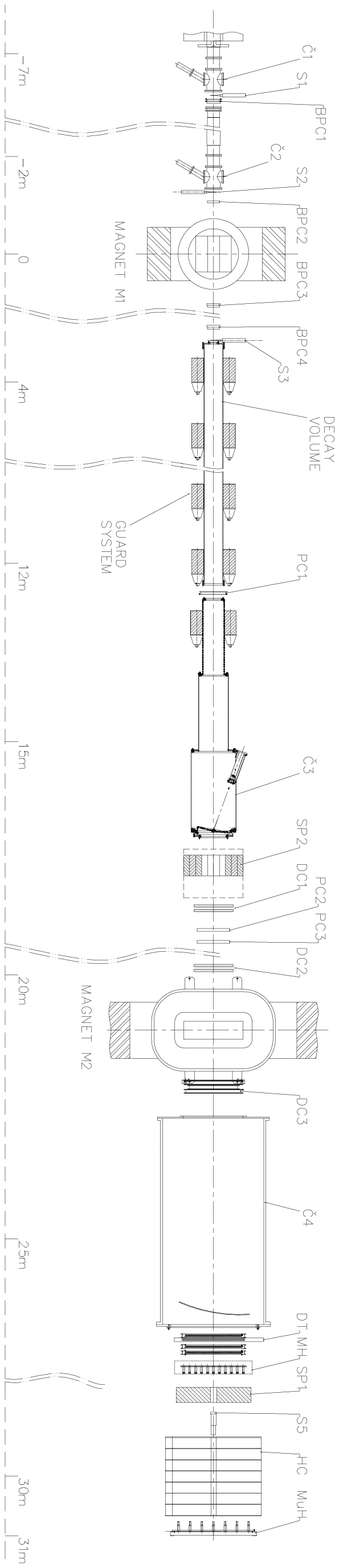,height=17cm,angle=90} 
\caption{Schematic view of the ISTRA+ setup.}
\end{figure}

 During  first run in March-April 2001, 363 million of
 trigger events were logged on DLT's.
 During second physics run in November-December 2001 350 million trigger
 events were collected with  higher beam intensity and somewhat stronger
 trigger requirements. This information
 is supported by about 100 million MC events generated using Geant3 \cite{ref5}
 for dominant $K^-$ decay modes. Signal efficiency for possible sgoldstino
  production has been estimated using one million generated
  events for the first run configuration
  and 0.5 million events for the second run configuration, for each effective
  mass $m_{P}$ point in the mass interval from 0 to 200~MeV with a step of 10~MeV.
  These signal events were weighted using the matrix element given in 
  \cite{ref1}.

\subsection{Event selection}

 Data collected in two  runs, Spring 2001, first run, 
 and Winter 2001, second run,  are used. Some information on the data
 processing and reconstruction procedures is given in \cite{ref2,ref3,ref4},
 here we briefly mention the details, relevant for 
$\pi^{-}\pi^{\circ}$~+~missing energy  event selection.

 The muon identification (see \cite{ref3}) is based on the information from the
  SP$_1$ --- a 576-cell lead glass calorimeter and the HC --- a scintillation-iron
 sampling calorimeter. The electron identification (see \cite{ref2})
 is done using $E/p$ ratio ---
 of the energy  of the shower associated with charged track and charged track
 momentum.
   
  A set of cuts is designed in order to suppress various
 backgrounds to possible sgoldstino production:

 0) Events  with one reconstructed charged track and with two
 electromagnetic showers  in the electromagnetic calorimeter SP$_1$ are selected.
 We require  the effective mass m$(\gamma\gamma)$ to be within
 $\pm 40$~MeV from m$_{\pi^{\circ}}$.
 The effective two-photon mass distributions 
\begin{figure}[tbh]
\epsfig{file=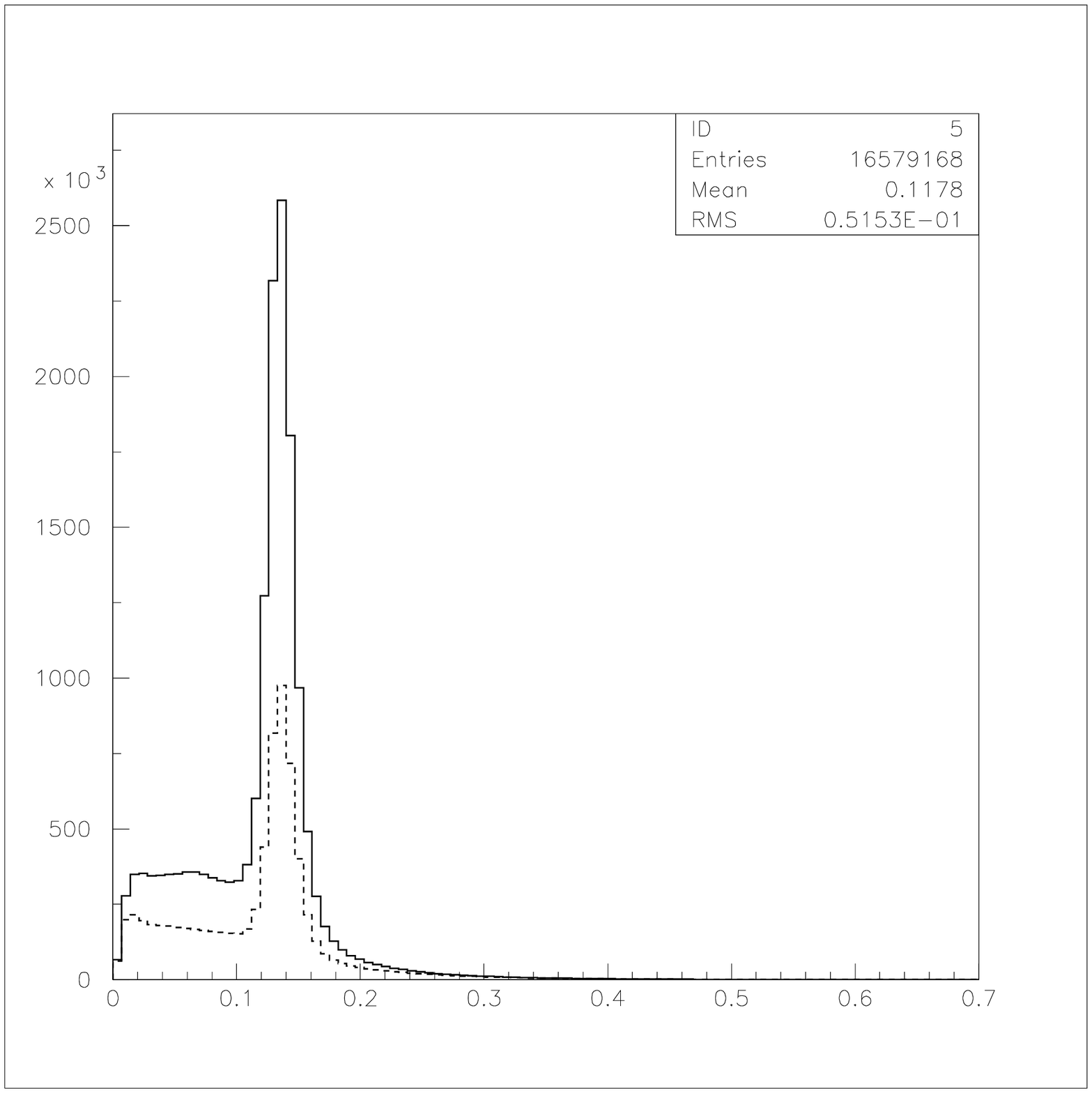,bbllx=31pt,bblly=154pt,bburx=550pt,bbury=647pt,%
width=16cm,clip=}
\caption{Two photon effective mass spectra  for events with one charged track
and two showers in the SP$_1$. Dashed line ---
 first run data, solid line --- second run data.}
\end{figure}
 for two runs are shown in Fig.~3.
  Events with vertex inside interval $ 400 < z < 1650$~cm are selected.

1) Next cut is a "soft" charged pion identification, tracks having
 electron or muon flag ( as described in \cite{ref2,ref3}~) are rejected.
 
2)   Events with  missing energy
 $E_{beam} - E_{\pi^{-}} - E_{\pi^{\circ}} $ below  3~GeV are rejected. 
 The cut on the missing
 energy serves to reduce $K_{\pi2}$ contribution. In Fig.~4 
missing energy spectra
  for the second run data, MC background and
 MC signal with $m_{P}$ of 90~MeV are compared. 
\begin{figure}[tbh]
\epsfig{file=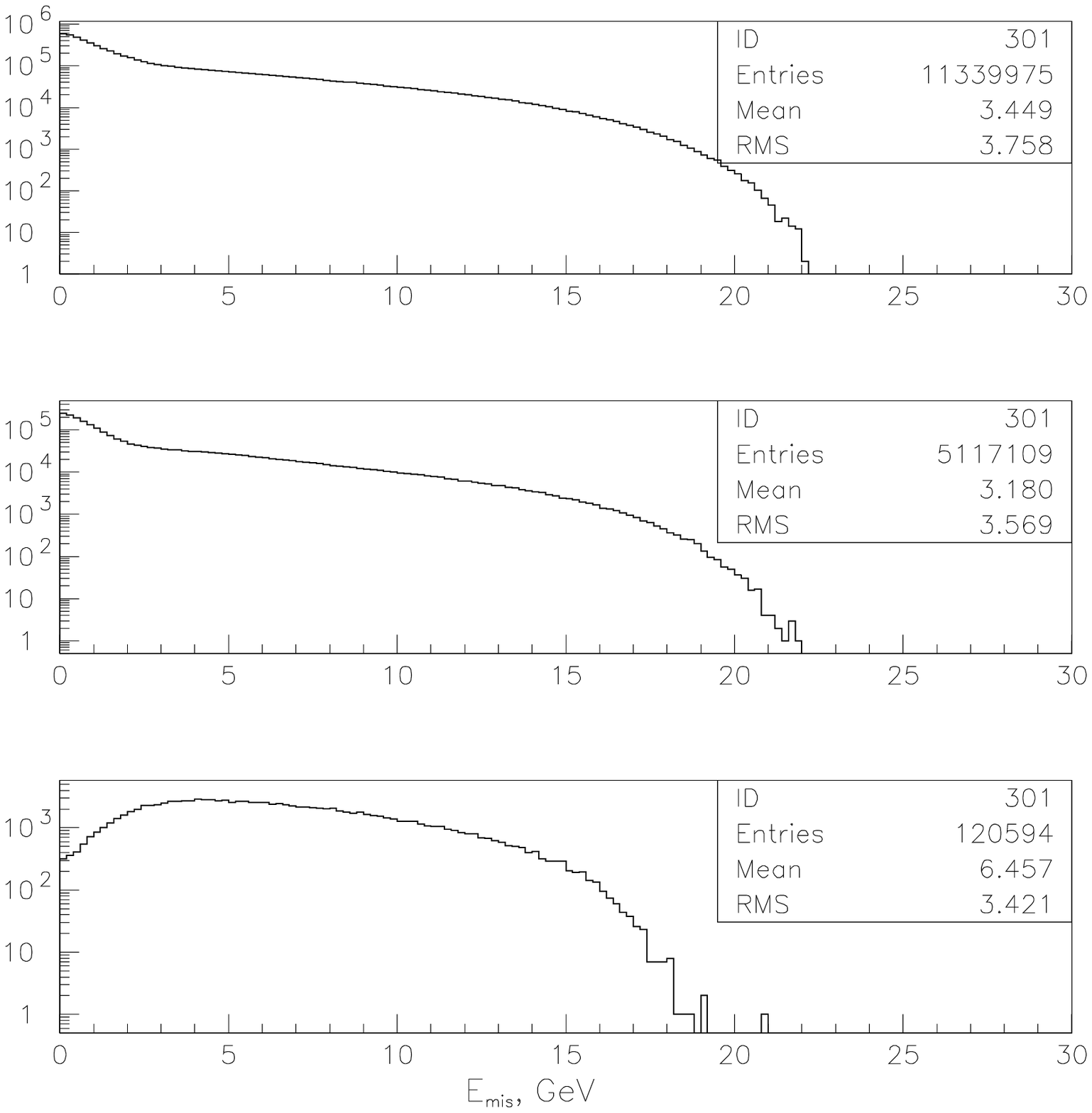,bbllx=31pt,bblly=144pt,bburx=550pt,bbury=683pt,%
width=16cm,clip=}
\caption{Missing energy spectra, second run data. Upper histogram ---
 real data, middle histogram --- background MC events, lower
 histogram --- signal MC events for $m_{P}=90$~MeV.}
\end{figure}

3)  Events with $m(\gamma\gamma)$ within $\pm 36$~MeV
 from m$_{\pi^{\circ}}$135~MeV a selected (Tough $\pi^{\circ}$ cut).

4) Fourth cut suppresses events with extra photons irradiated by charged particles 
 in a detector material upstream  M2-magnet.
 Such  photons have nearly the same $x$ coordinate on the SP$_1$ face
 as the charged track, i.e
 event is rejected if at least one photon has 
 $| x_{ch} - x_{\gamma} | < 10$~cm. 

5) Fifth cut removes events having good  $K_{e3}$ 2C-fit.

6) The following cut is a restriction on  the 
 charged pion momentum $p^*$ in the kaon
 rest frame: $p^*(\pi^-) < 180$~MeV. It is designed to suppress 
 the tails of the $K_{\pi2}$ decay.

7) Seventh cut is against remaining muon contamination.It is required that
 the energy in SP$_1$, associated with charged track, should exceed 0.7~GeV. 
 This cut requires "early" hadron shower in SP$_1$ and has , unfortunately,
 low signal efficiency. We plan to get rid of it in our final analysis.

8) The following cut is designed to suppress $\pi^{-}\pi^{\circ}\gamma$ contribution:
   only the events with missing momentum crossing SP$_1$ transverse plane within
   SP$_1$ working aperture are selected.
\begin{figure}
\epsfig{file=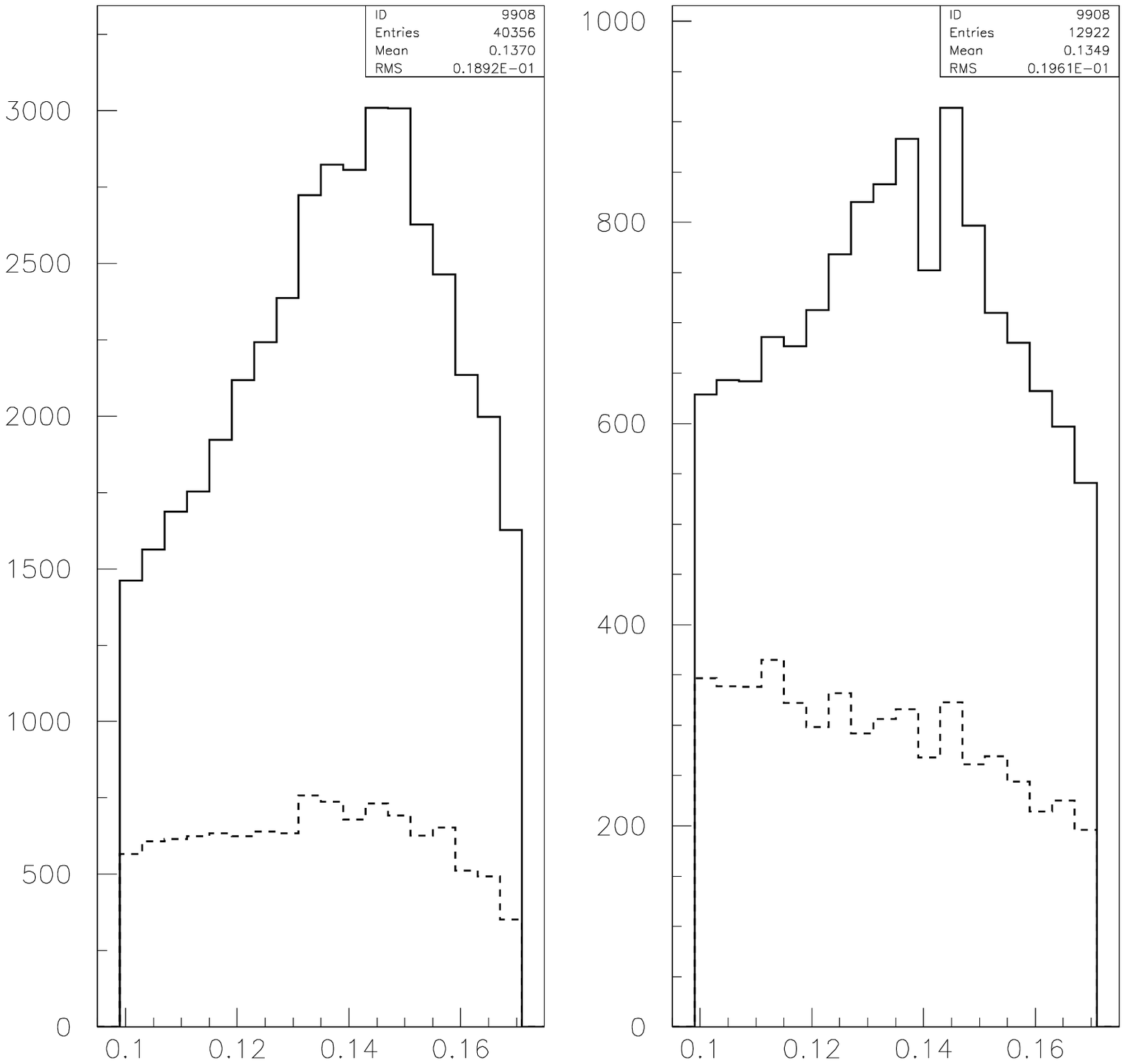,bbllx=31pt,bblly=154pt,bburx=550pt,bbury=683pt,%
width=16cm,clip=}
\caption{The influence of the "Veto" cut on the two-photon effective mass
 spectrum. Left part --- second run, right part --- first run; solid line ---
 before the "Veto" cut, dashed line --- after the "Veto" cut.}
\end{figure}

9) The  "Veto" cut uses information from the Guard System GS and the guard 
electromagnetic calorimeter SP$_2$: absence of  signals
above noise threshold is required.
 The influence of the "Veto" cut on the two-photon
effective mass spectrum is illustrated in Fig.~5. No $\pi ^{\circ}$
signal is seen after the "Veto" cut. 

10) As a result of the previous cuts, $\pi ^{\circ}$-signal 
practically disappears 
from the $\gamma \gamma$ mass spectrum. This allows to perform effective linear
background subtraction procedure. The interval  $\pm~ 18$~MeV around 
m$(\pi^{\circ})$ mass 
value is used as the signal region, remaining events are used 
for the background 
estimate.

The  data reduction information 
is given in Table~1 for two runs and is compared in Tables~2,3 with
 MC-background statistics  and  with
 MC-signal statistics for the sgoldstino mass of $m=90$~MeV.   

 The influence of the last eight cuts on the missing mass squared spectrum 
 $(P_{K}-P_{\pi^{-}}-P_{\pi^{\circ}})^{2}$, which is used as 
the signal "estimator" is 
 is shown in Fig.~6   for the second run data.
 The left wide bump in Fig.~6 is due to 
 $K_{\mu3}$  decays, the shift to negative missing mass squared 
  is caused by the use of the pion mass
 in its calculations. Second peak is caused by 
 $\pi^{-}\pi^{\circ}(\pi^{\circ})$ decay
 with gammas from second $\pi^{\circ}$ escaping detection.
 
 Fig.~7 illustrates the background subtraction procedure: 
 in the first(third) figure  the missing mass squared spectrum  
 after cut~10 is shown for
 the first(second) run (solid curve) together  with the normalized 
 background(dashed
 curve) estimated from the  $m(\gamma\gamma)$ spectrum side bands
 ( 99--117~MeV and 153--171~MeV). The second(forth) figure show 
signal-background
 difference for  two runs. No sgoldstino signal is observed, the peak around 
 0.02 GeV$^{2}$ is due to remaining $\pi^{-}\pi^{\circ}(\pi^{\circ})$ 
 background.

\begin{table}[tbhp]
\caption{ Event reduction statistics for two runs. } 
\begin{center}
\begin{tabular}{|c|c|c|c|c|}
\hline
 Cut  & Events run~1&   $N_i/N_{i+1}$ & Events run~2 &
 $N_i/N_{i+1}$    \\  
%\hline
% $N_{events}$  &  363.002.105 \\
\hline
(0) 1 $\pi^{-}$ and m$(\gamma\gamma)$ near m($\pi^{\circ}$) 
 & 4398982 & & 11339975 &  \\
\hline
(1)  no (e, $\mu$)  & 978752 & 1.48  & 8762220 & 1.29\\
\hline
(2) E$_{mis}>3.0$~GeV & 716830 &4.16 &  1430781 & 6.12\\
\hline
(3) narrow m($\gamma\gamma$) band  & 581653 & 1.23  & 1297087 & 1.10\\
\hline 
(4) conv. gammas & 322073 & 1.81  & 723248 & 1.79     \\
\hline
(5) no $K_{e3}$  fit & 230424 & 1.40 &  487246 & 1.48\\
\hline
(6) $p^*(\pi^-) <180$~MeV & 150926 & 1.53 & 322298 & 1.51\\
\hline
(7)$E_{SP_1} > 700$~MeV & 17101 & 8.83  & 51624 & 6.24\\
\hline
(8) $10<rr<60$~cm & 12922 & 1.32 & 40356 & 1.28 \\
\hline
(9) Veto & 5255 & 2.46 & 11183 & 3.61 \\
\hline
(10) $\pm 18$~meV & 2691 & 1.95 & 6125 & 1.94 \\
\hline
\end{tabular}
\end{center}
\end{table}
\renewcommand{\arraystretch}{1.0}

\begin{table}[t]
\caption{ Event reduction statistics for MC data, first run. } 
\begin{center}
\begin{tabular}{|c|c|c|c|c|}
\hline
 Cut  & Events mix MC&   $N_{i}/N_{i+1}$ & MC signal with m=90~MeV &
 $N_{i}/N_{i+1}$    \\  
%\hline
% $N_{events}$  &  363.002.105 \\
\hline
(0) 1 $\pi^{-}$ and m$(\gamma\gamma)$ near m($\pi^{\circ}$) 
 & 1278382 & & 207315 &  \\
\hline
(1)  no (e, $\mu$)  & 1156685 & 1.11  & 203841 & 1.02\\
\hline
(2) E$_{mis}>3.0$~GeV & 158181 &7.31 &  168179 & 1.21\\
\hline
(3) narrow m($\gamma\gamma$) band  & 149679 & 1.06  & 165105 & 1.02\\
\hline 
(4) conv. gammas & 91251 & 1.64  & 94503 & 1.75     \\
\hline
(5) no $K_{e3}$  fit & 48067 & 1.90 &  68325 & 1.38\\
\hline
(6) $p^*(\pi^-)<180$~MeV & 32151 & 1.50 & 67528 & 1.01\\
\hline
(7)$E_{SP_1} > 700$~MeV & 6880 & 4.67  & 22346 & 3.02\\
\hline
(8) $10<rr<60$~cm & 5283 & 1.30 & 17883 & 1.25 \\
\hline
(9) Veto & 2257 & 2.34 & 17770 & 1.01 \\
\hline
(10) $\pm 18$~MeV & 1521 & 1.48 & 15737 & 1.13 \\
\hline
\end{tabular}
\end{center}
\end{table}
\renewcommand{\arraystretch}{1.0}

\begin{table}[p]
\caption{ Event reduction statistics for MC data, second run. } 
\begin{center}
\begin{tabular}{|c|c|c|c|c|}
\hline
 Cut  & Events mix MC&   $N_{i}/N_{i+1}$ & MC signal with m=90~MeV &
 $N_{i}/N_{i+1}$    \\  
%\hline
% $N_{events}$  &  363.002.105 \\
\hline
(0) 1 $\pi^{-}$ and m$(\gamma\gamma)$ near m($\pi^{\circ}$) 
 & 5117109 & & 120594 &  \\
\hline
(1)  no (e, $\mu$)  & 4627156 & 1.11  & 119292 & 1.01\\
\hline
(2) E$_{mis}>3.0$~GeV & 618446 &7.48 &  98098 & 1.22\\
\hline
(3) narrow m($\gamma\gamma$) band  & 565772 & 1.09  & 96504 & 1.02\\
\hline 
(4) conv. gammas & 346226 & 1.63  & 54206 & 1.78     \\
\hline
(5) no $K_{e3}$  fit & 118536 & 2.92 &  37124 & 1.46\\
\hline
(6) $p^*(\pi^-)<180$~MeV & 114160 & 1.04 & 36835 & 1.01\\
\hline
(7)$E_{SP_1} > 700$~MeV & 21824 & 5.23  & 11800 & 3.12\\
\hline
(8) $10<rr<60$~cm & 16597 & 1.28 & 9462 & 1.25 \\
\hline
(9) Veto & 7229 & 2.30 & 9407 & 1.01 \\
\hline
(10) $\pm 18$~MeV & 5114 & 1.41 & 8541 & 1.10 \\
\hline
\end{tabular}
\end{center}
\end{table}
\renewcommand{\arraystretch}{1.0}

\begin{figure}[tbh]
%\begin{center}
\epsfig{file=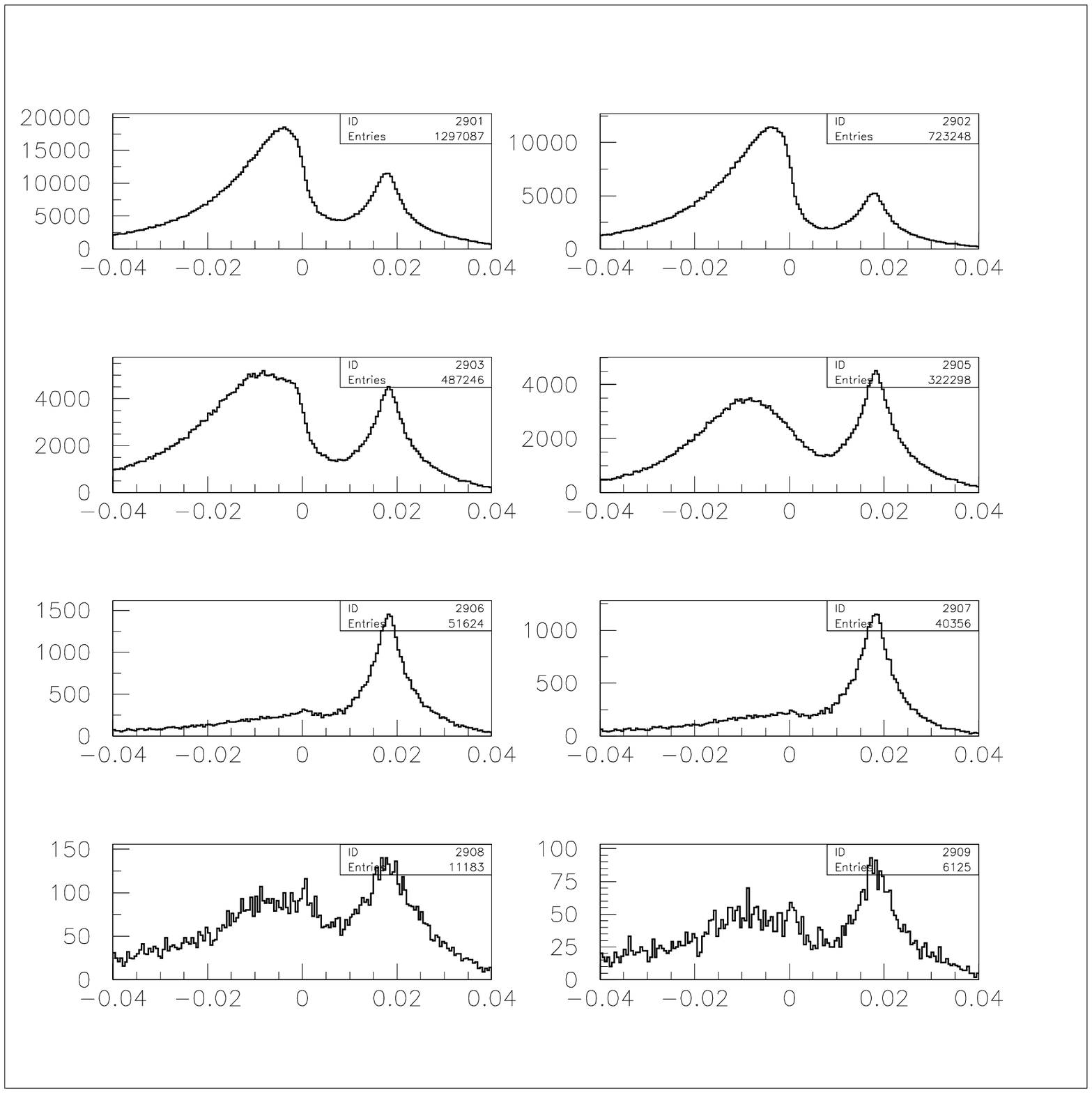,bbllx=31pt,bblly=154pt,bburx=550pt,bbury=647pt,%
width=16cm,clip=}
\caption{Missing mass squared distributions, real data, for successive
 cuts  3---10  respectively, second run data.}
%\end{center}
\end{figure}

\begin{figure}[tbh]
\begin{center}
\begin{tabular}{c}
\epsfig{file=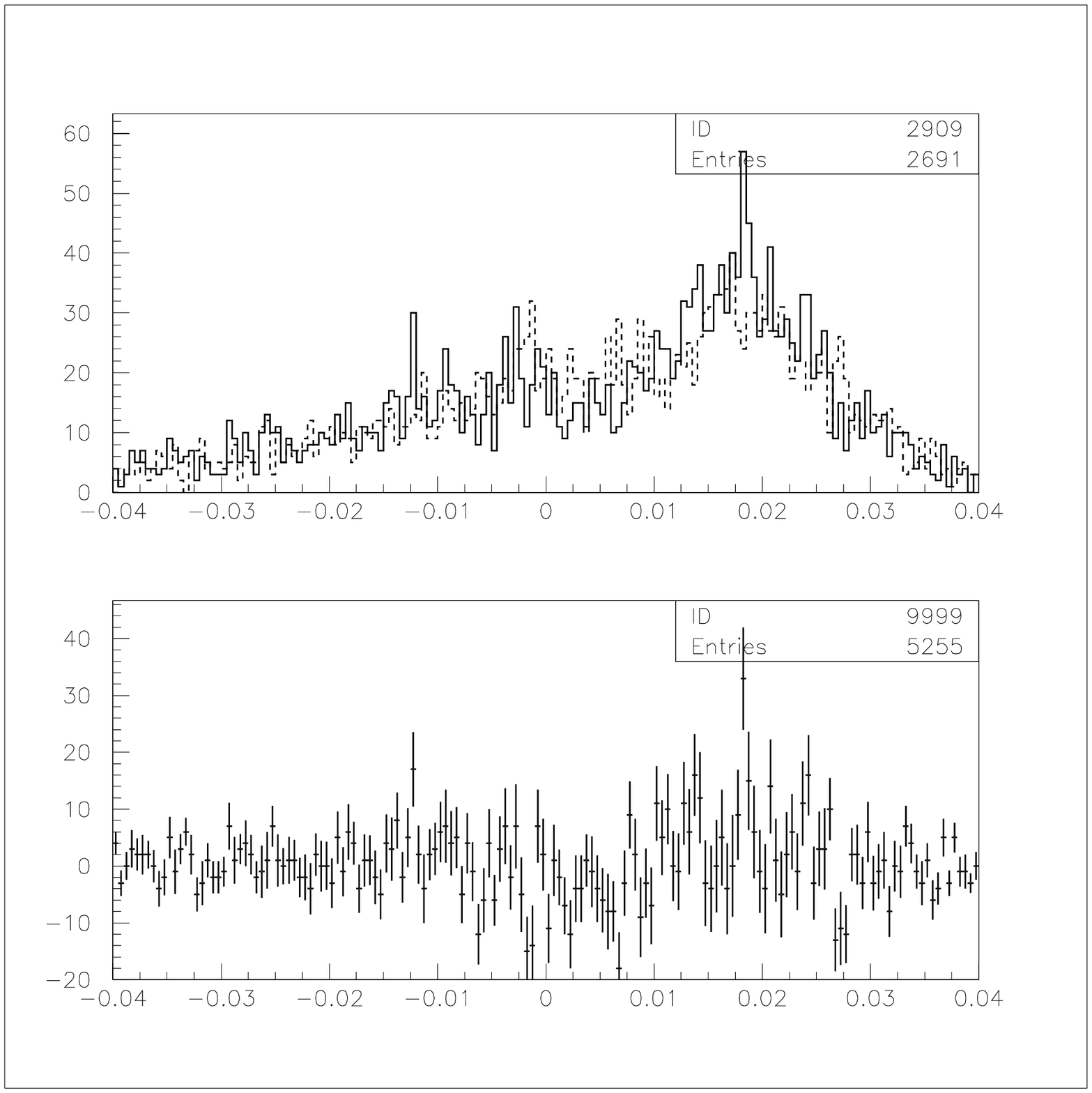,bbllx=31pt,bblly=155pt,bburx=550pt,bbury=646pt,%
height=9.0cm,clip=} \\
\epsfig{file=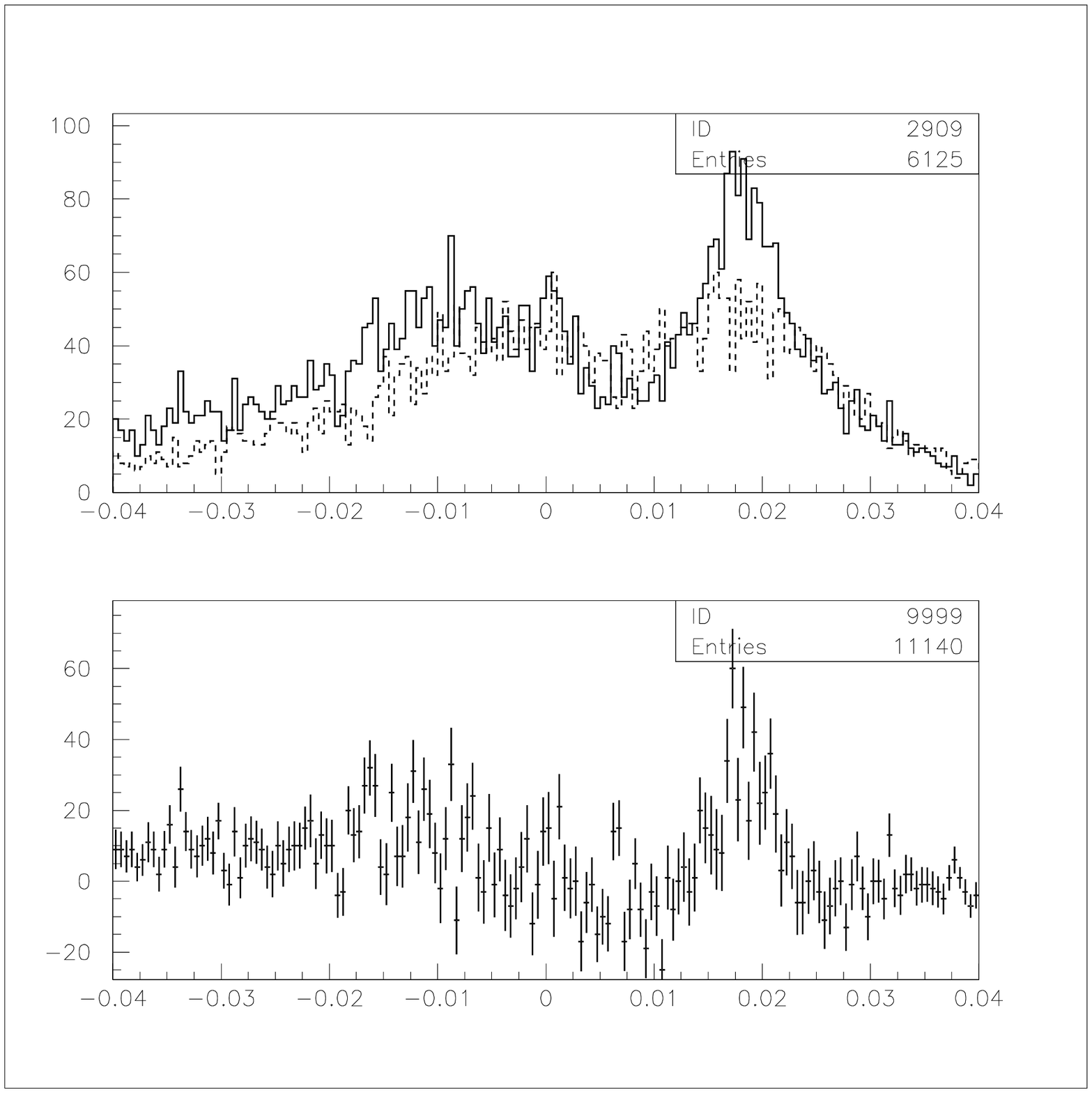,bbllx=31pt,bblly=155pt,bburx=550pt,bbury=646pt,%
height=9.0cm,clip=} \\
\end{tabular}
\end{center}
\caption{ Missing mass squared spectra, background subtraction, 
  first run --- upper part, second run --- lower part.  Solid line shows 
spectrum for
 the neutral pion selection within narrow interval,
 dotted line shows spectrum for
 tails. Lower histogram in each part shows the difference.}
\end{figure}

\section{ Analysis and results}

\begin{figure}[tbh]
\begin{center}
\begin{tabular}{c}
\epsfig{file=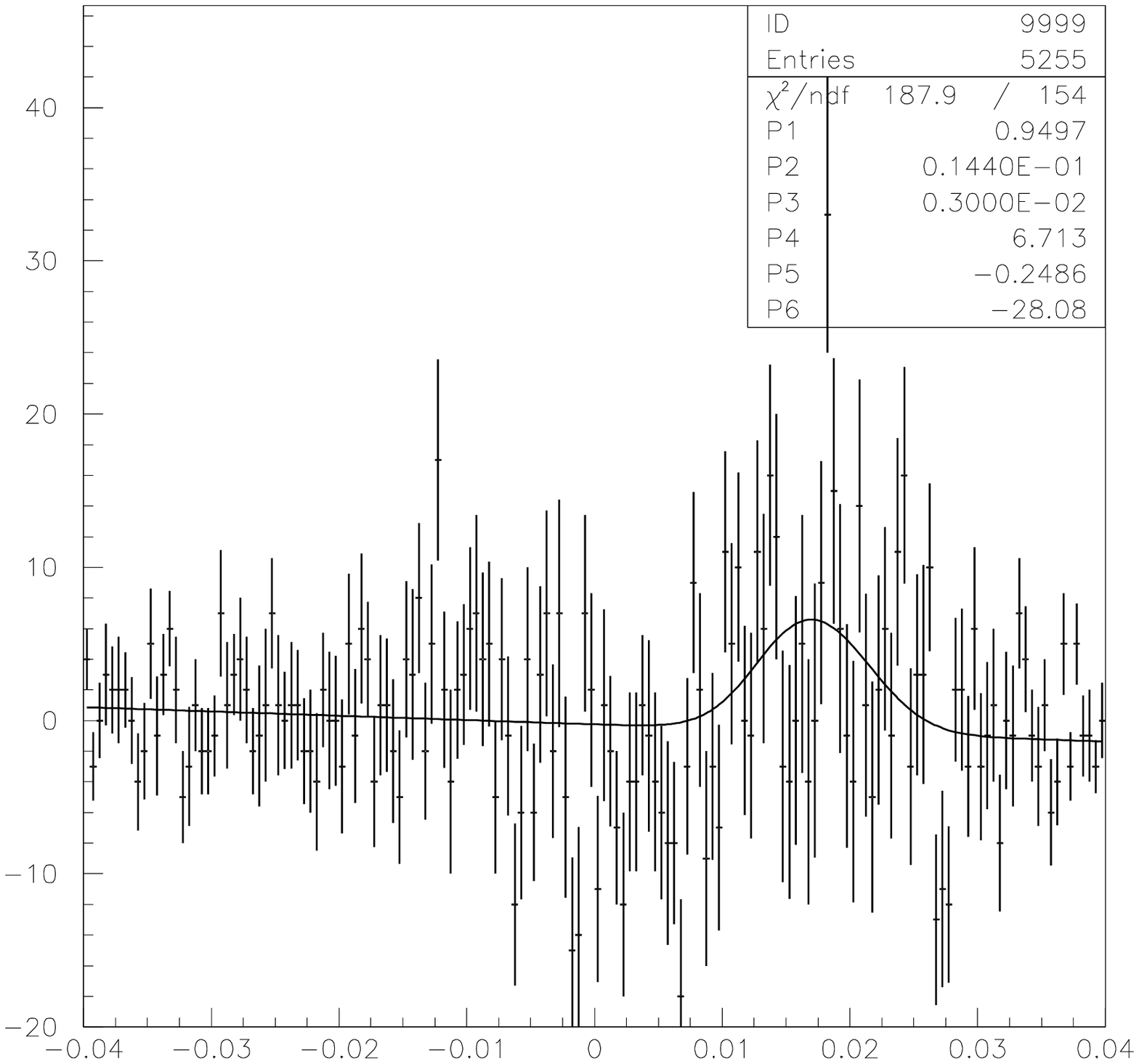,bbllx=38pt,bblly=180pt,bburx=547pt,bbury=645pt,%
height=9.0cm,clip=} \\
\epsfig{file=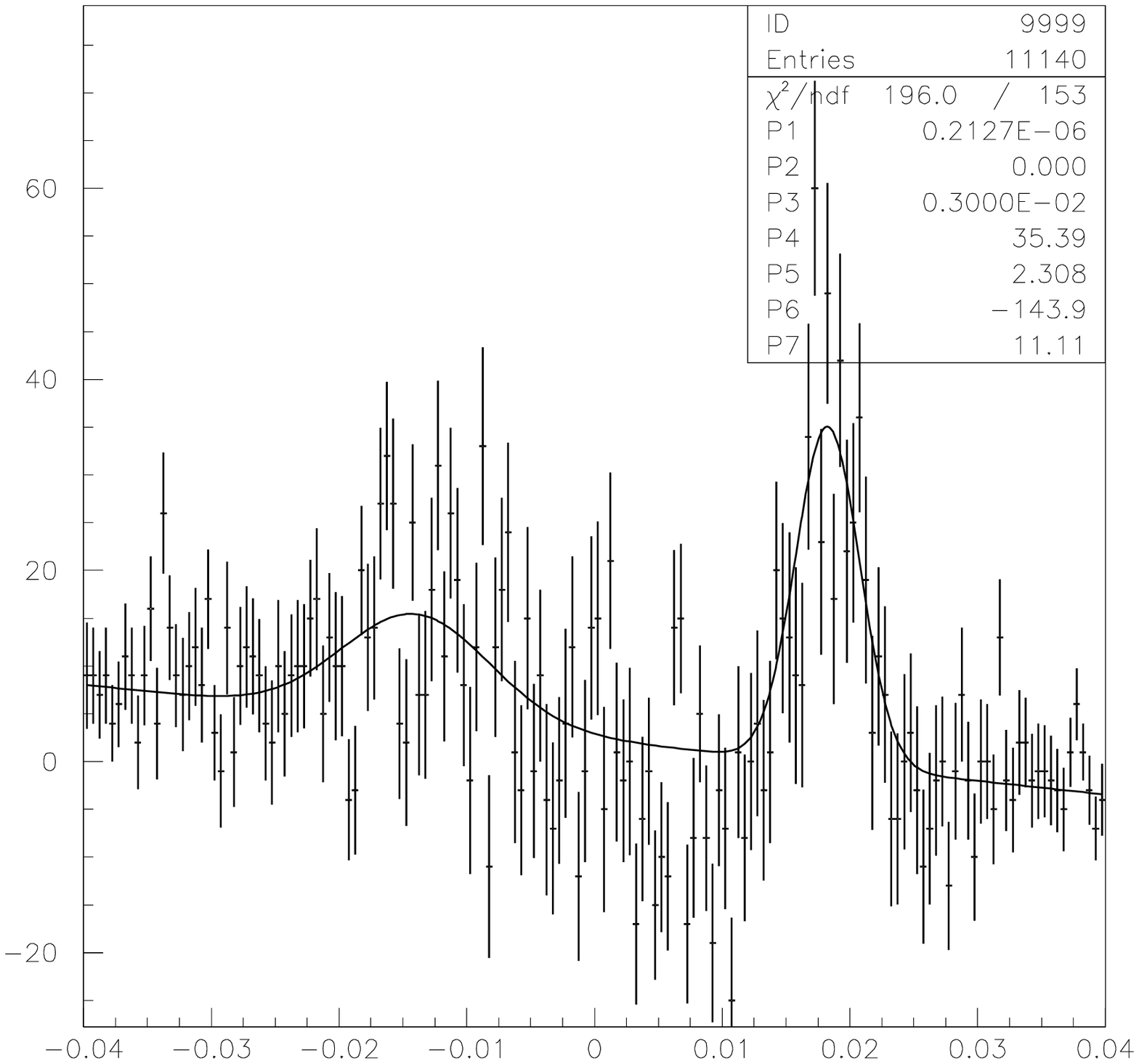,bbllx=38pt,bblly=180pt,bburx=547pt,bbury=645pt,%
height=9.0cm,clip=} \\
\end{tabular}
\end{center}
\caption{ Fit of  the missing mass squared spectrum  with
  sgoldstino signal at  120~MeV, first run data, upper part.
  Lower part --- fit of
 the missing mass squared spectrum with sgoldstino signal at 0~MeV,
 second run data. }
\end{figure} 

In order to calculate the upper limits
we have fitted the residual missing mass distribution by the sum of the 
background and the signal Gaussian with 
fixed width of $\sigma=0.003$~GeV$^2$, as 
determined from the signal MC. 

 The signal MC is also
 used  to determine the efficiency
of our cuts. The background  has been described by two components:
%, one Gaussian for remaining
% $K_{\mu 3}$ peak, 
the Gaussian for $\pi^-\pi^{\circ}(\pi^{\circ})$ peak
%, one wide Gaussian
% centered near -0.01 ( this component is explained by the pileup events) 
%and flat polynomial
%(namely quintic) background.  
 and a linear background. For the second run an additional  
wide Gaussian centered
 at -0.015 has been used. Fit results are illustrated in Fig.~7 for the first
 and second run, respectively. 
 
The upper limits at the
90$\%$ confidence level are calculated from the value
\begin{equation}
 N_{UL} = N_{obs} + 1.64 \cdot \sigma ~~~,
\end{equation}
  where $N_{obs}$ is the number of events in the signal  Gaussian.
 During the fit we impose the constraint that $N_{obs}$ is nonnegative. 
Sigma is a
 non-Gaussian error in $N_{obs}$. 
  The upper limit $UL$ is calculated as
  \begin{equation}
 UL =  
  \frac{N_{UL} \cdot 0.2116 \cdot 0.98798 \cdot \varepsilon(K_{\pi 2})}
 { N(K_{\pi 2}) \cdot \varepsilon}
\end{equation}
 with $0.2116$ equal to  $Br(K_{\pi 2})$ ;
 with 0.98798 equal to  $Br(\pi^{\circ} \rightarrow \gamma\gamma )$ and
 $N(K_{\pi 2})$ equal to the number of reconstructed $\pi^-\pi^{\circ}$ decays
 found to be 1500000 events for the first run and 4500000 events for the
 second run.

 The values $\varepsilon$ and $\varepsilon(K_{\pi 2})$ are respective
 acceptances for $K^{-} \rightarrow \pi^-\pi^{\circ}~P$ and $K_{\pi 2}$ decays.
 The acceptance $\varepsilon(K_{\pi 2})$ is equal to $17.49~\%$ for
 the first run and $24.92~\%$ for the second run. This acceptance includes
 both the reconstruction efficiency and geometrical acceptance.
% $N_{UL} = 1$.
 The results of the fits are given in Table~4. 
 As the background conditions for the two runs are different, 
 the first run provides
 better upper limits for some $m_{P}$ regions  
 regardless of the lower luminosity. The
 estimated  systematic errors for the upper limits are at the level of
 20~$\%$.  

\begin{figure}[tbh]
%\begin{center}
\epsfig{file=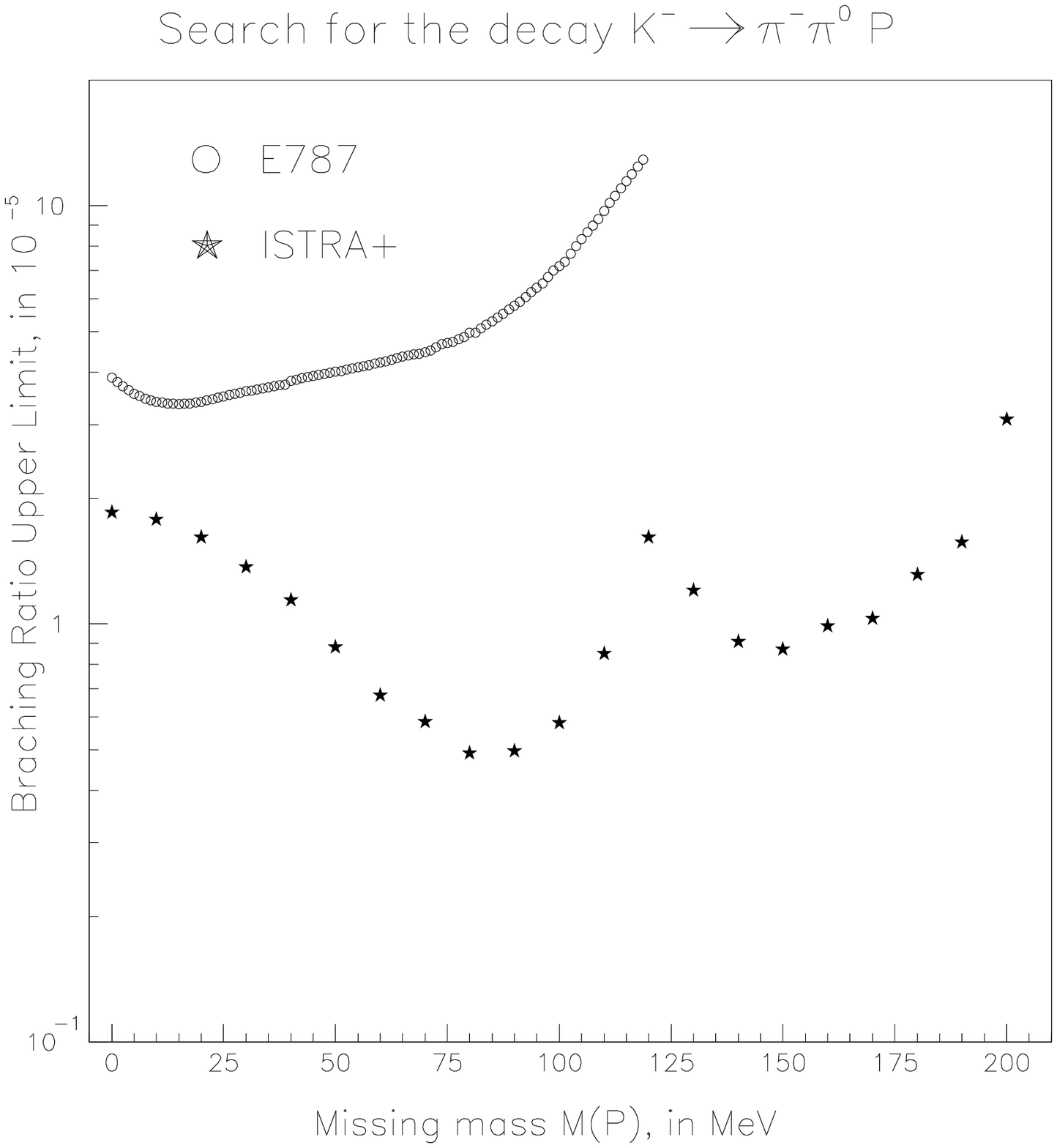,bbllx=28pt,bblly=128pt,bburx=558pt,bbury=705pt,%
width=16cm,clip=}
%\end{center}
\caption{ Mass dependence of the upper limits, calculated 
 using two runs together.}
\end{figure} 
 
\begin{table}
\caption{ Mass dependence of upper limits UL calculated using two 
 runs separately, left part --- first run, right part second run. } 
\begin{center}
\begin{tabular}{|c|c|c|c|c|c|c|}
\hline
 M, MeV  & Eff. $\varepsilon$, $\%$ & $N_{obs} \pm \sigma$ & 
%$N_{obs}+1.64~\sigma$ &
 UL & Eff. $\varepsilon$, $\%$& $N_{obs}\pm \sigma$& UL     \\  
\hline
 0 & 0.64 & 0$\pm 4.80$  & 1.84 $\cdot ~10^{-5}$  
  & 0.85& 0$\pm17.51$ & 2.39 $\cdot ~10^{-5}$\\
\hline
 10 & 0.66 & 0$\pm4.76$  & 1.76 $\cdot ~10^{-5}$ & 0.86 &0$\pm17.90$& 
 2.40 $\cdot ~10^{-5}$ \\                    
\hline
 20  & 0.67 & 0$\pm4.65$  & 1.69 $\cdot ~10^{-5}$ & 0.92 &0$\pm16.81$&
 2.10 $\cdot ~10^{-5}$  \\
\hline 
 30 & 0.74 & 0$\pm4.49$  & 1.48 $\cdot ~10^{-5}$ & 0.99 &0$\pm14.88$& 
 1.73 $\cdot ~10^{-5}$      \\
\hline
 40 & 0.81 & 0$\pm4.32$  & 1.30 $\cdot ~10^{-5}$ &1.05&0$\pm12.32$& 
 1.65 $\cdot ~10^{-5}$  \\
\hline
 50  & 0.88 & 0$\pm4.19$  & 1.16 $\cdot ~10^{-5}$ &1.18&0$\pm9.73$& 
 1.56 $\cdot ~10^{-5}$ \\
\hline
60 &  1.05  & 0$\pm4.19$  & 0.98 $\cdot ~10^{-5}$ &1.32&0$\pm7.67$& 
1.10 $\cdot ~10^{-5}$\\
%\hline
%65 & 1.09 & 0$\pm4.51$  & 1.01 $\cdot ~10^{-5}$ & 4.4 $\cdot
% ~10^{-5}$ \\
\hline
70 & 1.09 & 0$\pm4.51$  & 1.01 $\cdot ~10^{-5}$ &1.73&0$\pm6.35$& 
0.70 $\cdot ~10^{-5}$ \\
\hline
80 & 1.21 & 0$\pm5.71$  & 1.15 $\cdot ~10^{-5}$ &1.62&0$\pm5.62$& 
0.66 $\cdot ~10^{-5}$ \\
\hline
90 & 1.33 & 0$\pm10.97$  & 2.02 $\cdot ~10^{-5}$ &1.76&0$\pm5.26$& 
0.57 $\cdot ~10^{-5}$ \\
\hline
100 & 1.41 & 0$\pm24.59$  & 4.24 $\cdot ~10^{-5}$ &1.90&0$\pm5.40$& 
0.54 $\cdot ~10^{-5}$ \\
\hline
110 & 1.53 & 10.96$\pm18.16$  & 3.96 $\cdot ~10^{-5}$ &2.09&0$\pm6.86$& 
0.62 $\cdot ~10^{-5}$ \\
\hline
120 & 1.60 &17.22$\pm15.99$ &  2.24 $\cdot ~10^{-5}$ &2.27&0$\pm12.44$& 
1.04 $\cdot ~10^{-5}$ \\
\hline
130 & 1.63 & 0$\pm25.86$    & 3.86 $\cdot ~10^{-5}$ & 
      2.36 & 0$\pm28.13$    & 2.26 $\cdot ~10^{-5}$\\
\hline
140 & 1.66 & 0$\pm25.72$ & 3.79 $\cdot ~10^{-5}$ & 
      2.46 & 0$\pm27.42$ & 2.12 $\cdot ~10^{-5}$\\
\hline
150 & 1.61 & 8.27$\pm22.97$ &4.25 $\cdot ~10^{-5}$ & 
      2.49 & 0   $\pm24.85$ &1.89 $\cdot ~10^{-5}$ \\
\hline
160 & 1.74 & 0$\pm$14.91 & 2.33 $\cdot ~10^{-5}$ & 
      2.41 & 0$\pm$20.96 & 1.65 $\cdot ~10^{-5}$ \\
\hline
170 & 1.48 & 0$\pm$20.95 & 3.46 $\cdot ~10^{-5}$ & 
      2.35 & 0$\pm$19.98 & 1.61 $\cdot ~10^{-5}$ \\
\hline
180 & 1.42 & 0$\pm$21.00 & 3.60 $\cdot ~10^{-5}$ & 
      2.18 & 15.0$\pm13.35$& 1.61 $\cdot ~10^{-5}$\\
\hline
190 & 1.27 & 0$\pm$16.19 & 3.11 $\cdot ~10^{-5}$ & 
      2.01 & 0$\pm$29.29 & 2.77 $\cdot ~10^{-5}$\\
\hline
200 & 0.85 & 0$\pm$15.89 & 4.54 $\cdot ~10^{-5}$ & 
      1.26 & 6.49$\pm$15.96 & 3.00 $\cdot ~10^{-5}$\\
\hline
\end{tabular}
\end{center}
\end{table}
\begin{table}
\caption{ Mass dependence of upper limits UL calculated using two runs
 together. } 
\begin{center}
\begin{tabular}{|c|c|c|}
\hline
 M, MeV   & $N_{obs} \pm \sigma$ & 
%$N_{obs}+1.64~\sigma$ &
 UL       \\  
\hline
 0  & 0$\pm 11.21$  & 1.85 $\cdot ~10^{-5}$  \\
\hline
 10  & 0$\pm11.05$  & 1.78 $\cdot ~10^{-5}$  \\                    
\hline
 20   & 0$\pm10.56$  & 1.61 $\cdot ~10^{-5}$   \\
\hline 
 30  & 0$\pm9.72$  & 1.37 $\cdot ~10^{-5}$       \\
\hline
 40  & 0$\pm8.62$  & 1.14 $\cdot ~10^{-5}$   \\
\hline
 50   & 0$\pm7.42$  & 0.88 $\cdot ~10^{-5}$  \\
\hline
60   & 0$\pm6.39$  & 0.68 $\cdot ~10^{-5}$   \\
%\hline
%65 & *6.25 & 0$\pm6.35$  & *1.01 $\cdot ~10^{-5}$ & 4.4 $\cdot
% ~10^{-5 \\
\hline
70  & 0$\pm5.78$  & 0.58 $\cdot ~10^{-5}$  \\
\hline
80  & 0$\pm5.69$  & 0.49 $\cdot ~10^{-5}$  \\
\hline
90  & 0$\pm6.27$  & 0.50 $\cdot ~10^{-5}$  \\
\hline
100  & 0$\pm7.88$  & 0.58 $\cdot ~10^{-5}$  \\
\hline
110  & 0$\pm12.59$  & 0.85 $\cdot ~10^{-5}$  \\
\hline
120  & 0$\pm25.67$ &  1.61 $\cdot ~10^{-5}$  \\
\hline
130  & 0$\pm19.91$    & 1.20 $\cdot ~10^{-5}$  \\
\hline
140  & 0$\pm15.51$ & 0.91 $\cdot ~10^{-5}$  \\
\hline
150  & 0$\pm14.94$ & 0.87 $\cdot ~10^{-5}$  \\
\hline
160  & 0$\pm16.79$ & 0.99 $\cdot ~10^{-5}$  \\
\hline
170 & 0$\pm17.07$ & 1.03 $\cdot ~10^{-5}$  \\
\hline
180 & 0$\pm20.09$ & 1.31 $\cdot ~10^{-5}$  \\
\hline
190 & 0$\pm22.12$ & 1.57 $\cdot ~10^{-5}$  \\
\hline
200 & 13.66$\pm19.06$& 3.08 $\cdot ~10^{-5}$  \\
\hline
\end{tabular}
\end{center}
\end{table}

   The final result 
  obtained using combined statistics of two runs  is given
  in Table~5. The weighted ratio 
  $\varepsilon(K_{\pi 2})/\varepsilon$ with the weights proportional to
  the runs statistics has been used  for the combined data sample.
    Fig.~9 shows a comparison of our result 
   with that published by the E787 collaboration \cite{ref6}.
  
 Special attention should be paid to the region near m$_{\pi^{\circ}}$ mass.  
 If we assume that the sgoldstino mass  is exactly equal 
to that of $\pi^{\circ}$
 and that  the  remaining $\pi^{\circ}$ peak is due to sgoldstino signal,
 then the upper limit for this point is  $2.8 \cdot ~10^{-4}$.

\section{Summary and conclusions}

 A search for a possible pseudoscalar sgoldstino production in the decay 
 $K^{-} \rightarrow \pi^{-} \pi^{\circ} P$ has been performed. 
It was assumed that
 sgoldstino decays outside the setup, i.e is invisible.
 No signal is seen in the $m_{P}$ mass interval between 0 and 200~MeV.
 The 90$\%$ confidence upper limits obtained vary
 between $2.0 \cdot ~10^{-5}$ and $0.5 \cdot ~10^{-5}$ 
for the sgoldstino mass range
 of $0 \div 190$~MeV. These results  improve the confidence limits
 published by the E787 Collaboration.  
  
  In future, we plan to extend our search for a sgoldstino signal
  in the scenario when  the decay $P\rightarrow \gamma\gamma$ 
happens inside the setup.

The authors would like  to thank D.S.~Gorbunov, V.A.~Matveev, V.A.~Rubakov, 
for  numerous discussions. 

The INR part of the collaboration 
is  supported by the RFFI fund contract N00-02-16074.

\end{document}